\documentclass{article}
\usepackage{amssymb,amsmath,graphicx,float}
\usepackage{cite,citesort}
\newcommand{\gref}[1]{(\ref{#1})}
\newcommand{\td}{\text{d}}

\newcommand{\de}[2]{\frac{\text{d} #1}{\text{d}#2}}
\newcommand{\dde}[2]{\frac{\text{d}^2 #1}{\text{d} #2 ^2}}
\newcommand{\be}{\begin{equation}}
\newcommand{\ee}{\end{equation}}

\title{Stochastic `fuzzy confinement' of intrinsic localised modes}
\author{Matthias Meister\footnote{email: matthias@unizar.es}
\\
Dpto. F\'\i sica de la Materia Condensada, Facultad de Ciencias, and\\
Instituto de Biocomputaci\'on y F\'\i sica de
Sistemas Complejos,\\ Universidad de
Zaragoza, 50009 Zaragoza, Spain \\ \quad \\
L. V\'azquez\\
Dpto. Matem\'atica Aplicada, Facultad de Inform\'atica, \\Universidad Complutense de
Madrid, 28040 Madrid, Spain, and \\
Centro de Astrobiolog\'\i a (CSIC-INTA),\\
28850 Torrej\'on de Ardoz, Spain}
\begin{document}
\maketitle
\begin{abstract}
The long time diffusive behaviour of intrinsic localised modes (discrete breathers) in the discrete
damped-driven sine-Gordon chain under Gaussian white noise (to simulate temperature) is studied. We
present a theoretical model for an approximate description of the diffusion, derive an expression
for the diffusion constant and compare with results from simulations. It turns out that an increase
of the temperature inhibits the diffusive motion in such a way that the breather, propagating with a
well-defined velocity in the noise-free case, is, with increasing
probability, confined to a finite region of the system.
As all physical processes in
the real world occur at temperatures $T>0$, these results also have a bearing on the
experimental detection of mobile breathers, e.g. in parallel arrays of
Josephson junctions.
\end{abstract}
\begin{figure}
\begin{center}
\includegraphics[scale=0.7]{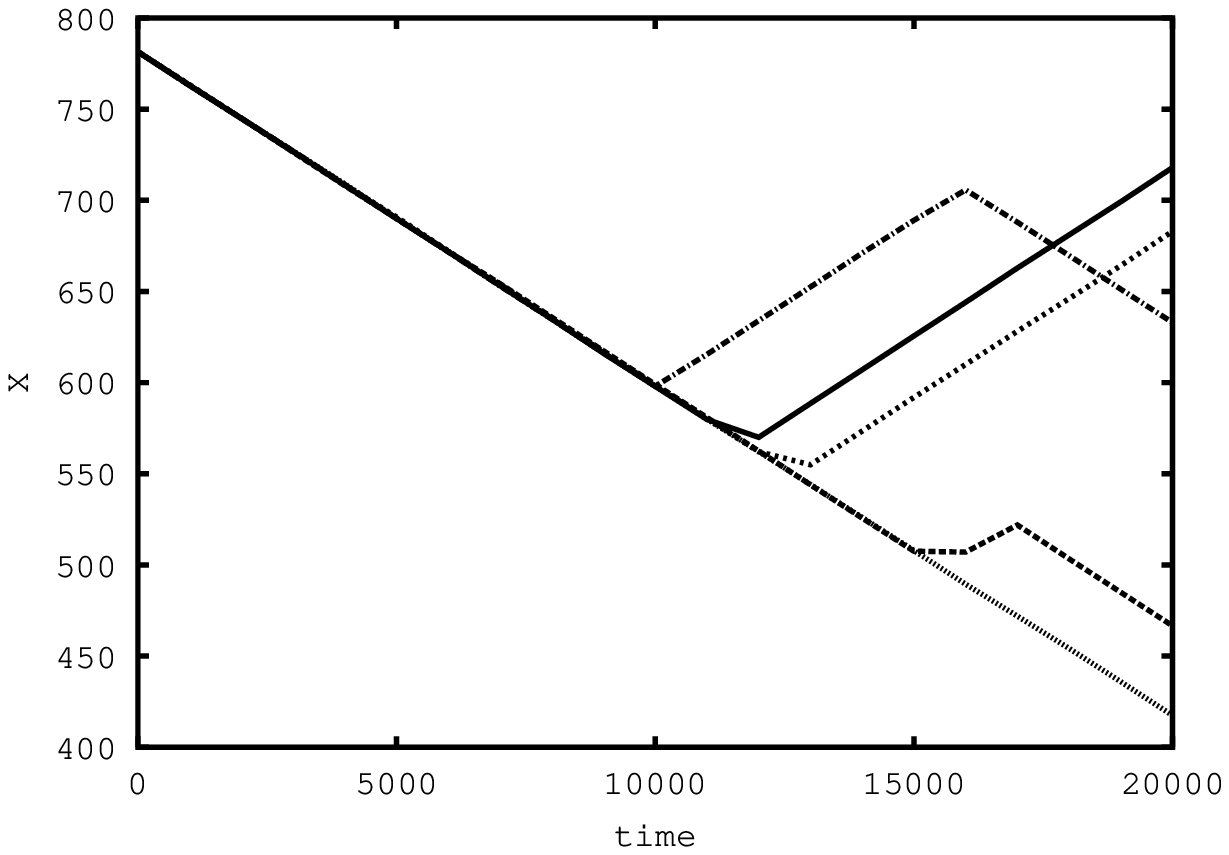}\\
\includegraphics[scale=0.7]{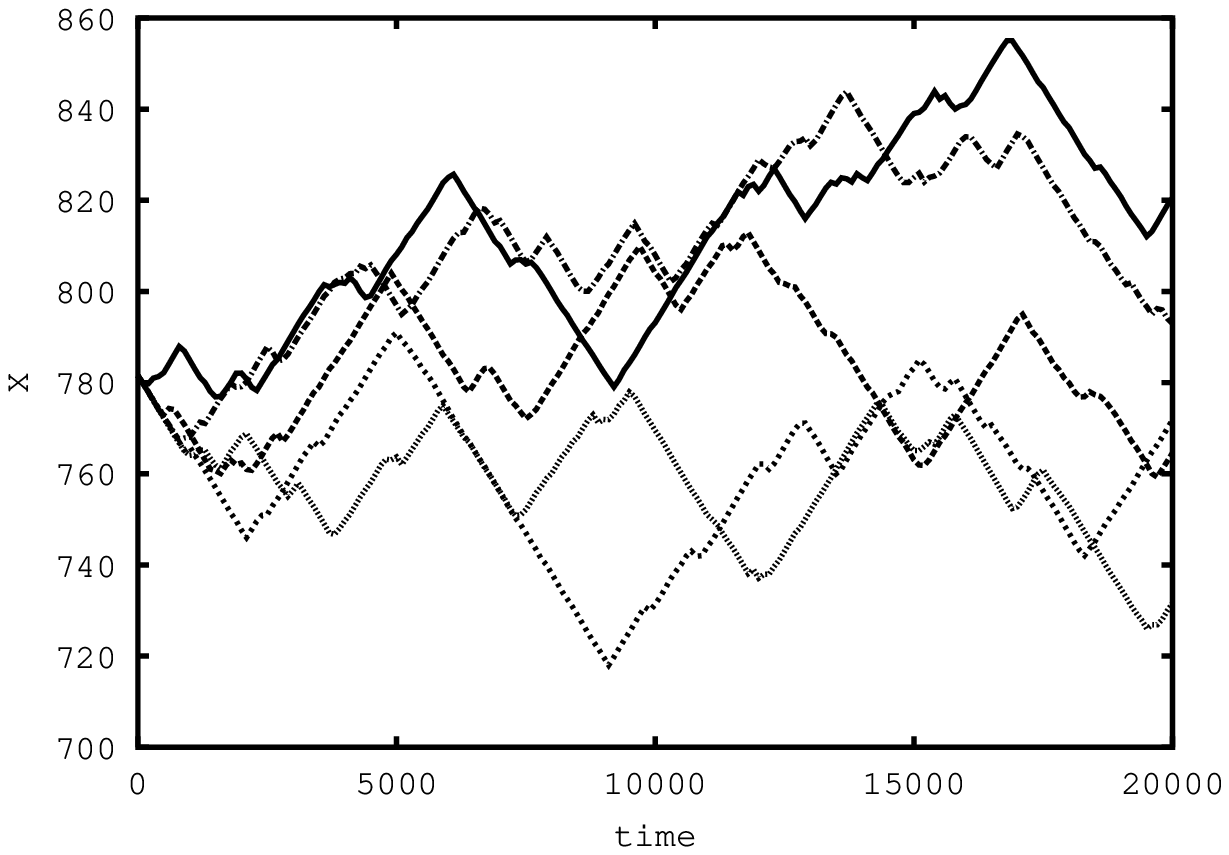}
\end{center}
\caption{The position $X$ of a discrete breather as a function of time
for several realisations of the noise. Top: $\sigma=5\cdot 10^{-5}$,
corresponding to $k_{B}T=6.25 \cdot 10^{-8}$, bottom: $\sigma=11\cdot 10^{-5}$, corresponding to
$k_{B}T=3.025 \cdot 10^{-7}$.
Clearly an increase of the temperature increases the number of velocity flips.}
\label{Trajectories}
\end{figure}

Intrinsic localised modes (ILMs) are localised excitations that can exist even in perfectly
homogeneous discrete nonlinear systems due to an interplay of discreteness and nonlinearity.
These modes, also referred to as discrete breathers (DBs), have received much attention over
the past 10-15 years.
At least in part this is due to the fact that little is needed for their existence, as shown by
existence proofs \cite{MacKay94,Aubry97,Sep97} for this type of localised
excitation. Therefore DBs are rather generic and can occur for example in magnetic
lattices \cite{Kha03,Schw99}, nonlinear optical waveguide systems \cite{Per03,McGurn03}, Josephson
junction arrays \cite{Ust03,Maz03} or nonlinear oscillator networks, e.g. \cite{Mar01,Mart03}: thus
DBs are of broad interest. More extensive references to the literature, covering both theoretical
and experimental aspects, can be found in the cited publications.
\newline
We study the discrete
sine-Gordon equation under damping and driving, 
\be \label{model}
\dde{u_{n}}{t}=-\frac{1}{2\pi}\sin\left(2\pi u_{n}\right)-\alpha \de{u_{n}}{t} +
F\sin(\omega_{0}t)+\\C\left(u_{n+1}+u_{n-1}-2u_{n}\right)+\sigma \xi_{n}
\ee
which is already written in terms of dimensionless quantities.
This model describes a chain of pendula in a homogeneous gravitational field, neighbouring pendula
being harmonically coupled.
Here $u_{n}$ is the angle of deviation from the equilibrium position in units of
$2\pi$, $\alpha$ measures the strength of the damping, $F\sin(\omega_{0}t)$ is a
harmonic external driving, $C$ gives the strength of the coupling between
nearest-neighbour sites and the $\xi_{n}$ are Gaussian white noise terms of vanishing
mean and $\langle \xi_{n}(t)\xi_{m}(t^{\prime}) \rangle=\delta_{mn}\delta(t-t^{\prime})$,
$\sigma$ scaling the strength of the noise. 
The noise represents thermal fluctuations, the relation between
$\sigma$ and the temperature $T$ 
given by $\sigma^{2}=2\alpha k_{B}T$. A more appropriate  treatment of
the white noise could be achieved in terms of increments of Wiener
processes \cite{Gard}, 
which also would avoid the $\delta$-singularity in its correlation.
We do, however, not use the formalism of stochastic differential equations in this article, therefore
we do not deem it necessary to introduce these expressions here.
Equation \gref{model} can for instance also model a parallel array of Josephson junctions
\cite{Maz03}, in which case $u_{n}$ is the phase difference between the two superconductors in the
$n$-th junction, given in units of $2\pi$ and the driving force corresponds to a bias current. In
\cite{Mar01,Mart03} mobile intrinsic localised modes have been found to exist in model \gref{model}
(without the noise term) as attractors of the dynamics in certain
regions of parameter space. Correspondingly, the velocity of the mobile
ILMs depends on the system parameters ($C,\:\alpha,\:\omega,\:F$),
but, after a transient, not on the precise initial conditions. At some
values of the parameters more than one basin of attraction
exists, each corresponding to a different value of the velocity. In all cases there is a degeneracy
of the model due to symmetry, as for a given velocity $v$ there are always basins of attraction
corresponding to $+v$ and $-v$. The diffusive behaviour of the localised excitation under the
influence of the white noise $\sigma \xi$ can be described, as we will show in this article, by a
hopping of the system between these two degenerate basins of attraction. At the parameter values we
are using in the simulations for this article, $\alpha=0.02, F=0.02,
\omega_{0}=0.2\pi, C=0.890$ (a choice which to our knowledge is
generic within the parameter regime where mobile ILMs exist), there are known three different basins
of attraction with different (absolute) values of the velocities. 
We solve equation \gref{model} numerically using the Heun-algorithm,
with a time step $\Delta t = 0.01$, which corresponds to $0.001 \times 2\pi/\omega_{0}$.
It turns out in simulations that
transitions between basins corresponding to different velocities do not play a role in the process,
so we focus only on one value of the velocity, in our case $v\approx 0.0186$.
\newline
A noise induced flip of the velocity, i.e. a transition from one attracting configuration to a
different one, involves a probably complex dynamical reconfiguration of the system. Unable at
present to exactly describe this process, we aim at approximately capturing the main features of the
transition in terms of a simple model. We assume that for fixed parameters there is a constant
probability $\Theta$ per unit time for a transition to be excited. At least in the ranges of
parameter space we have explored numerically, we have found the plausible result that the transition
probability increases with the noise strength (temperature), see figure \ref{Trajectories}.
For the sake of completeness we add here the comment that breathers
also can be destroyed or created by thermal fluctuations, or form out
of generic energy distributions in a system; in the
temperature range and on the time scales we have used here, such
processes have not been observed, and therefore will not play a role
in this work. Related discussions can be found for instance in \cite{Pey98,Pey00}. 
We have to take into account that the
reconfiguration accompanying a velocity flip is not an instantaneous process. Rather than
considering details of this process we introduce a time scale $\tau_{A}$, which can be considered
the (average) duration of the reconfiguration. We model the transition process in the following way:
With probability $\Theta$ per time unit a transition is initiated. If this occurs, in our simplified
description the velocity of the ILM immediately assumes the value $0$ and the breather remains
immobile for a time span $\tau_{A}$; this drop of the velocity to $v=0$ suggested itself as the most
tractable approximation to the behaviour of $v$ in a transition that flips the sign of the velocity.
After the delay $\tau_{A}$ the velocity assumes the value opposite to its initial value. As during
the `waiting time' $\tau_{A}$ the system is already involved in reconfiguration, we do not allow any
further jumps to be initiated during the waiting time. If therefore a jump has been initiated at
time $t_{i}$ the probability for {\it no further jump} to occur at least up to time
$t$ is
\be
1-H(t-t_{i}-\tau_{A})+H(t-t_{i}-\tau_{A})\exp\left[-\Theta(t-t_{i}-\tau_{A})\right]
\ee
where $H$ is Heaviside's step function.
The probabilities for $n$ jumps in time $t$ can also be derived and enter in the calculation of the
expectation values and variances of the position, which involve rather
tedious algebra and can be found in \cite{JPA}. Here we are only interested in the long time behaviour,
which can be obtained in a much simpler way. Long time in this context means
observation times $t \gg \frac{1}{\Theta}$, the latter being the mean time
before a jump of the velocity once the delay period $\tau_{A}$ has passed; note
that for the parameter values we have been looking at our numerical results have shown that
$\tau_{A} < \Theta^{-1}$. Thus we will be considering the case of stochastic trajectories
with many jumps in the observation time $t$.
Suppose that a jump has been initiated at $t=0$ and a second jump will be initiated
at some time $\tau_{A}<t^{\prime}<t$. Then the mean distance the ILM travels between the jumps
is
\be
\langle \Delta X \rangle = v\Theta\int \limits_{\tau_{A}}^{t}
(t^{\prime}-\tau_{A})e^{-\Theta(t^{\prime}-\tau_{A})}\td t^{\prime} \approx \frac{v}{\Theta}
\ee
Here we have taken into account that the velocity is zero and jumps are forbidden for the time
interval $\tau_{A}$ after the jump at $t=0$; also we have made use of the long time condition
$t \gg \Theta^{-1}$ to obtain the approximate equality. Furthermore we find
\be
\langle \Delta X^{2} \rangle =
v^{2}\Theta\int\limits_{\tau_{A}}^{t}(t^{\prime}-\tau_{A})^{2}e^{-\Theta(t^{\prime}-\tau_{A})}
\td t^{\prime} \approx 2\frac{v^{2}}{\Theta^{2}}
\ee
so that we get
\be
\text{Var}\left(\Delta X\right)\approx \frac{v^{2}}{\Theta^{2}}
\ee
Introducing the parity $\pi(n)$ of a natural number $n$, i.e. $\pi(n)=+1$ if $n$
is even and $\pi(n)=-1$ if $n$ is odd, we can write for the distance
travelled  after $N$ jumps:
\be
\label{Njumps}
X_{N}=-\sum\limits_{i=1}^{N} \pi(i)\Delta X_{i}
\ee
where $\Delta X_{i}$ is the length of the $i$-th segment of the path of the
discrete breather, i.e. the segment terminated by the $i$-th jump. In \gref{Njumps} we
obviously start with an ILM of positive velocity.
We straightforwardly obtain
\be
\langle X_{N} \rangle=\frac{1-\pi(N)}{2} \frac{v}{\Theta}
\ee
and
\be
\text{Var}(X_{N})=\left[N-\frac{1-\pi(N)}{2}\right]\frac{v^{2}}{\Theta^{2}}
\ee
We are more interested in the dependence of the mean value and variance on time
than on the number of jumps. The mean number of jumps after time $t$
is
\be
\langle N \rangle = \frac{t}{\tau_{A}+\frac{1}{\Theta}}
\ee
because the average time between subsequent jumps is the sum of the waiting time $\tau_{A}$
and the mean time $\Theta^{-1}$ the system stays in one attractor after the passage of the waiting
time. As we consider the long time limit, we have $\langle N \rangle \gg 1$, and therefore the
relative weight of trajectories with an even number of jumps and those with an
odd number of jumps will be approximately equal. In this way we arrive at the
following results:
\be
\label{Mean}
\langle X(t) \rangle =  \frac{v}{2\Theta}
\ee
and
\be
\label{Var}
\text{Var}\left[X(t)\right]=\left[
\frac{t}{\tau_{A}+\frac{1}{\Theta}}-\frac{1}{2}\right] \frac{v^{2}}{\Theta^{2}}
\ee
the latter giving a diffusion constant $D=v^{2}/[\Theta^{2}(\tau_A+\Theta^{-1})]$.
The nonvanishing mean in \gref{Mean} is due to the asymmetry introduced
by the choice of the initial condition in \gref{Njumps}. When the
average numbers of realisations in the positive and negative velocity
attractors have equilibrated, the mean position does not change any
more. 
\begin{figure}
\begin{center}
\includegraphics[scale=0.7]{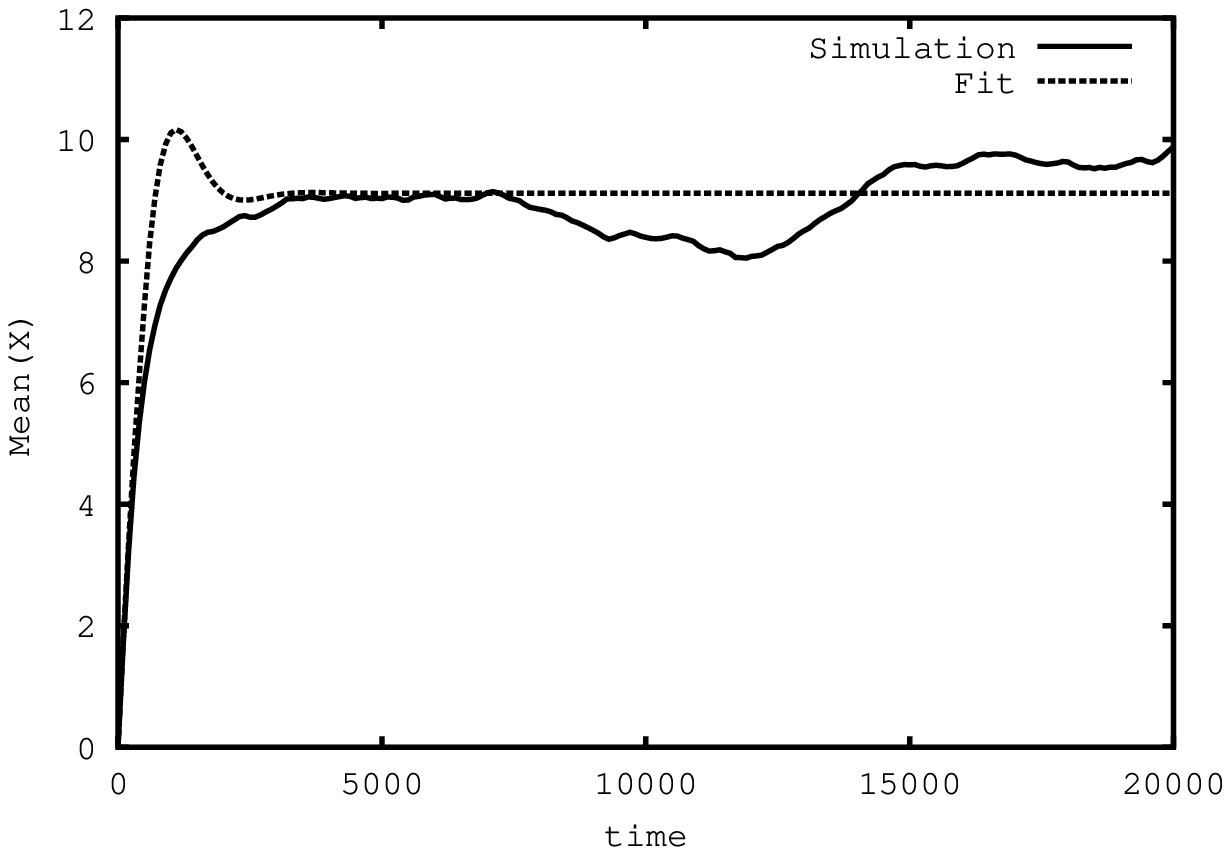}\\
\includegraphics[scale=0.7]{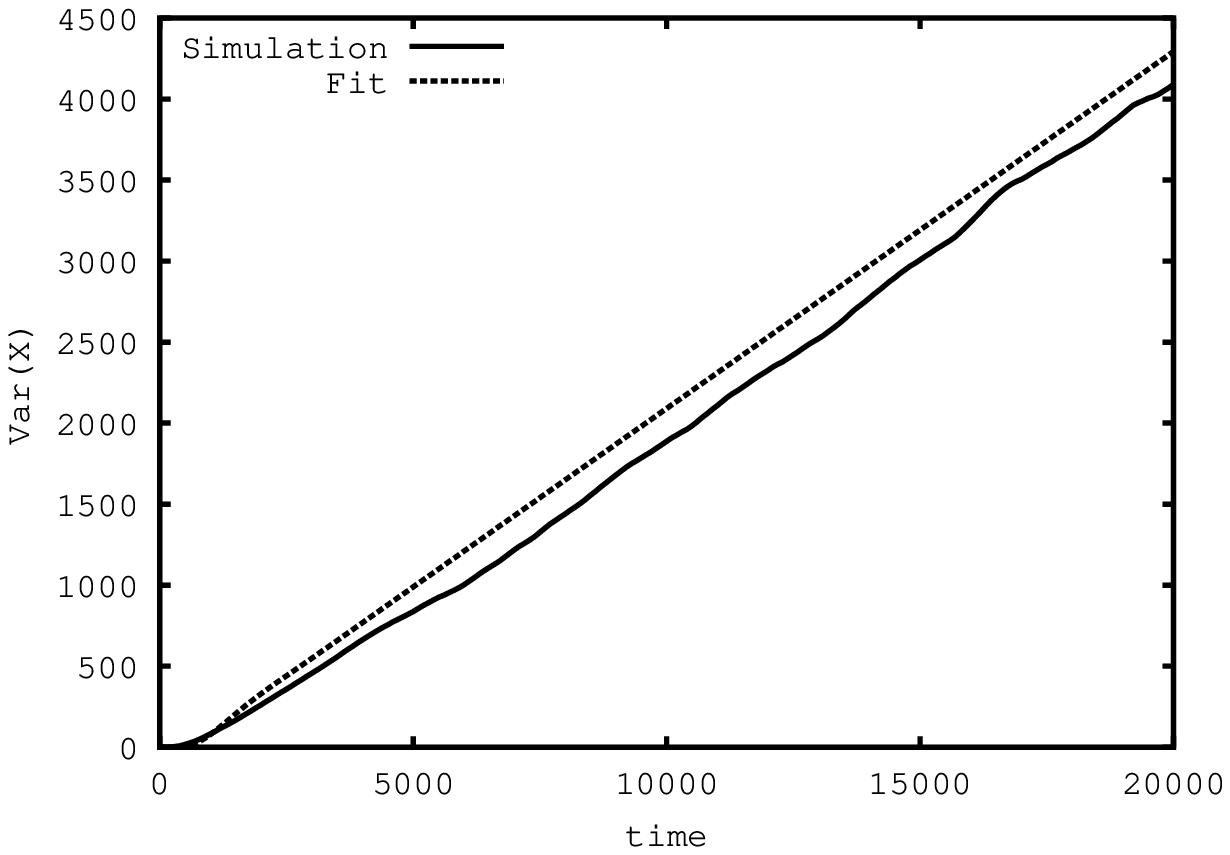}
\end{center}
\caption{Comparison of the predictions of the model with simulations. Top: mean value of the
position, bottom: variance; the simulation results are averages over
2000 realisations of the noise ($\sigma=12\cdot 10^{-5}$,
corresponding to $k_{B}T=3.6\cdot 10^{-7}$), starting from one and the
same initial configuration.
The fitted lines correspond to predictions from the full model, the parameters $\Theta=1.019\cdot
10^{-3}$ and $\tau_{A}=527.57$ were derived from the behaviour of the numerical results over the
final 10000 time units and equations \gref{Mean} and \gref{Var}. See
main text for details. The overshooting of the theoretical result for the mean
value of the position is due to the crudeness of the model, as
discussed further in the text.}
\label{Results}
\end{figure}
 In figure \ref{Results} these results are compared with
simulations. The numerical results for the mean value and the variance
of the position have been obtained as averages over 2000 realisations.
Of course there still occur fluctuations in these numerical
averages. In order to obtain a definitive value for the mean of the
position to be used in \gref{Mean} we average the numerically obtained
mean value over the final 10000 time units (which
can be considered to be in the `long time regime') and use this averaged value in equation
\gref{Mean} to calculate $\Theta$. Likewise, the mean slope of the numerically obtained variance over the
final 10000 time units and the calculated value of $\Theta$ are introduced in equation \gref{Var} to obtain
$\tau_{A}$. The two parameters are then used in the full model
\cite{JPA} to
generate a prediction for the mean and the variance of the position.
For the mean value we see from figure \ref{Results} that there is a distinct deviation between the
numerical and analytical results in the transition range between short time and long time behaviour.
This deviation occurs because of the crudeness of our model and would disappear if
the details of the reconfiguration during a velocity flip were taken
into account more fully. Numerical results show that this overshooting
increases, if the ratio $\tau_{A}/\Theta^{-1}$ increases, i.e. if the
time the configuration spends in transition increases relative to the
time the system is in one of the attracting states. This can easily be
understood: Our model does not take into account the details of the
transitions between the attractors, so here the description is only
approximate. Consequently, if the system spends more time in
transition, our description gets less accurate. 
A
closer look at the transition region reveals that the observed
deviation is also responsible for
the shift between the numerical result and the prediction for the 
variance in the
\begin{figure}[t]
\begin{center}
\includegraphics[scale=0.7]{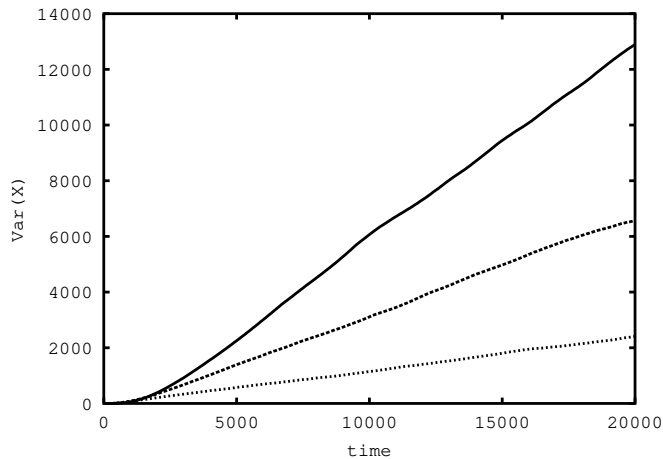}
\end{center}
\caption{The variance of the position of an ILM as a function of time for three different values of
the noise strength (results from simulations, averages over 2000 realisations). From top to bottom
the lines correspond to: $\sigma=8\cdot 10^{-5},k_{B}T=1.6\cdot 10^{-7};\sigma=1\cdot
10^{-4},k_{B}T=2.5\cdot 10^{-7}; \sigma=1.5\cdot 10^{-4},k_{B}T=5.625\cdot 10^{-7}$. In the
long time regime we observe a decrease of the slope of the variance with increasing temperature
(noise strength).} \label{Variances} \end{figure}
\begin{figure}[t]
\begin{center}
\includegraphics[scale=0.7]{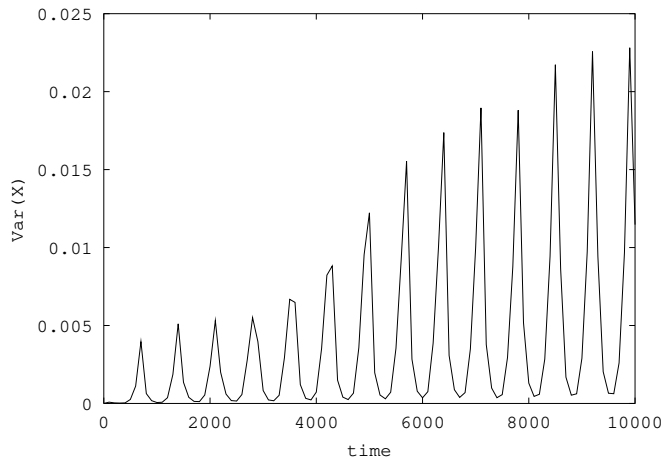}
\end{center}
\caption{
The variance of the position of an ILM at $\sigma=10^{-5}$
$(k_{B}T=2.5\cdot 10^{-9})$ obtained from averaging over 100
realisations.
In these realisations not a single jump occured. The distance between
the peaks does {\it not} correspond to the time the breather needs to
travel between neighbouring lattice sites (this is about 50
time units). Nonetheless the peaks are
most probably a discreteness effect:
As the breather is travelling along the system, it experiences a Peierls-Nabarro like potential, which, 
due to the internal dynamics of the breather, does not only depend on position, but on time, too. This time-dependent 
potential can amplify the effects of noise, increasing a noise-induced position difference between ILMs 
of different realisations by driving them into different adjacent cells. This should happen when the breathers cross 
borders between lattice cells and at the same time pass through an adequate internal configuration. A 
temporary increase of the variance should be observed. Once the retarded breather crosses into the cell of the 
advanced one, the PN potential may drive the retarded ILM towards the advanced, reducing now their positional 
separation it increased before.}
\label{LowNoise}
\end{figure}
long time regime. The slope of the variance in the 
long time regime is well reproduced by the full model, as is the
long time behaviour of the mean value. \newline
We see from equation \gref{Var} that an increase in the temperature (meaning also an increase in
$\Theta$) reduces the slope of the variance as a function of time, which has been observed in
simulations as well, see figure \ref{Variances}. Furthermore this result is also plausible, because
an increase in $\Theta$ implies an increase in the mean number of jumps that occur along a
trajectory within a given time. Thus the distance a breather can travel away from its average
position before the next flip of the velocity is reduced. In average for high $\Theta$, the ILM is
therefore confined to a small region of the system, where it is erratically moving back and forth.
Of course, this confinement is
not strict, but an issue of probabilities. With increasing time the probability that an ILM exits
from a fixed part of the chain grows, but the higher the temperature the slower this growth is.
We would like to point out that small noise is sufficient to effect this behaviour; all noise
strengths $\sigma$ for which we have shown results in this article are below one percent of the
driving amplitude $F$. 
This said, it may look as if by reducing the noise strength $\sigma$ to
arbitrarily low values the slope of the variance of $X$, as appearing
for example in figure \ref{Variances}, could be increased without
limit. We have not observed this, instead we found behaviour like in figure \ref{LowNoise}. However, our observation
times necessarily have been of limited extension, whereas our
results here apply to the long time case, i.e. 
a large average number of jumps should occur within the observation time.
By reducing $\sigma$ the transition probability between the
attracting configurations of the system becomes smaller and smaller and
we reach the point where, on average, only a few velocity flips occur
within the limited observation time. In this case the long time
condition is violated, and our results no longer apply. What would
happen if we were to increase the observation time (and necessarily
also the number of realisations) by several orders of magnitude, while
lowering the noise strength to significantly smaller values than the
ones we have used, is a question we deem unsuitable for our computational resources. Will 
there eventually appear a long time behaviour as for higher noise values or will discreteness 
effects like in figure \ref{LowNoise} dominate the dynamics? 
\newline
In summary we have shown that the presence even
of small thermal noise can change the behaviour of a mobile intrinsic localised mode in the
damped-driven sine-Gordon chain from deterministic translation with a well defined velocity to
propagation determined by random flips of the sign of the velocity. With increasing temperature the
number of flips increases and the mode is stochastically confined to a part of the system of
decreasing size. These results clearly have to be taken into account in any experiments that aim at
the detection of propagating ILMs in a system which can be described by the damped-driven sine-Gordon
model, like for instance a parallel array of Josephson junctions under an ac bias current.
As our description of the diffusion and qualitatively also the observed phenomenon itself
depend only on the existence of attracting configurations
corresponding to propagation velocities $+v,-v$ of a localised mode,
we expect analogous behaviour of propagating discrete breathers in
other systems which show these characteristics.
\newline\quad\newline
MM and LV acknowledge support from the European Commission within the Research Training
Network (RTN) LOCNET, contract HPRN-CT-1999-00163. LV also was partially supported by the
Ministry of Science and Technology of Spain through grant BFM2002-02359.

\end{document}